\documentstyle{aa}
\begin{document}
\thesaurus{03 (13.07.2; 11.01.2; 11.17.4 PKS0528+134)}
\title{COMPTEL Observation of the Flaring Quasar PKS0528+134}
\author{S. Zhang$^{1,2}$, T.P. Li$^1$, and M. Wu$^1$}
\institute {High Energy Astrophysics Laboratory,
Institute of High Energy Physics, Chinese Academy of Sciences,  Beijing
(zhangs@astrosv1.ihep.ac.cn)
\and
Institute of Theoretical Physics, Chinese Academy of Sciences, Beijing}
\titlerunning{COMPTEL Observation of the Flaring Quasar PKS0528+134}
\authorrunning{S. Zhang et al.}
\date{Received 10 July 1998 / Accepted 11 September 1998}
\maketitle
\begin{abstract}
With a direct demodulation  
method, we have reanalyzed the data from the COMPTEL observation of 
PKS0528+134 during the 1993 March flare in $\gamma$-rays. 
Our results show that
during the flare $\gamma$-rays were detected at a level approximately 2.4-3.8
times higher than the observed intensity in two earlier COMPTEL observations
VP 0 and VP 1 in the energy range 3 MeV to 30 MeV. The 3-30 MeV time variability 
of the flux
follows well the trend as observed by EGRET at higher energies. No 
convincing excess can be found around the position of PKS0528+134
in the energy range 0.75 MeV to 3 MeV, which indicates a spectral break 
around 3 MeV. The detections and non-detections in the four standard
COMPTEL energy bands are consistent with the earlier reports given by Collmar
et al.,
while the feature that $\gamma$-rays of the quasar still kept on flaring 
at energies down to 3 MeV is clearly found. 

\end{abstract}

\section{Introduction}

    The quasar PKS0528+134 was first detected at $\gamma$-ray energies by
EGRET aboard the Compton Gamma-Ray Observatory ({ \sl CGRO })(Hunter et al. 1993).
Before this, it had been known as a bright radio source with a flat radio
spectrum, a optical source with a mean brightness of $m_v$=20 and little 
polarization (Fugmann \& Meisenheimer 1988) and also an X-ray source (Bregman 
et al. 1985). It has a redshift of $z$=2.07 (Hunter at al. 1993) and, along 
with the strong $\gamma$-ray sources Crab and Geminga, is located near the 
Galactic anticenter. Following the detection of PKS0528+134 by EGRET, 
the COMPTEL data were analyzed and PKS0528+134 was discovered at soft 
$\gamma$-rays (0.75-30 MeV) by Collmar et al. (1993a). 
Collmar et al. (1993b) also reported the preliminary result from 
COMPTEL viewing periods 0 and 1 that spectral breaks were found at energies 
around 10 MeV 
when comparing with the spectra detected by EGRET. PKS0528+134
was seen to flare during viewing period (VP) 213 by EGRET and 
$\gamma$-rays ($\geq$ 100 MeV) were 
detected at a level approximately three times higher than the observed 
intensity in earlier observations VP 0 and VP 1 (Mukherjee et al. 1996).
The quasar was also detected by EGRET in its 'low' states
during the other observations include viewing periods 2.1, 36.0
, 36.5, 39, 221, 310, 321, and 337 (Nolan et al. 1993).
Comparisons of
the results from COMPTEL VP 213 with those from EGRET were carried out by Collmar et al.
(1997) in their analyses of PKS0528+134 using the first 3.5 yr of COMPTEL 
observational data (VP 0 to VP 337). They subdivided the data into 
two parts according to the luminosity of the source: viewing periods 0, 1 
and 213 as 
high states, and all the other observations as low states. They found
that during the low states the combined spectra can be fitted by a single 
power-law representation, while a spectral break at
energies around 10 MeV exists for the combined spectra of the high states.
However, their results for the flux show that PKS0528+134 had high luminosity 
in the 3-10 MeV band and, at the upper COMPTEL energy band (10-30 MeV), 
the quasar only had normal 
luminosity during VP 213. This feature is odd
and stimulates us to reanalyze the flaring quasar.

Considering the poor statistics of the data from COMPTEL VP 213 which covered
only 6 days, we adopt the direct demodulation method which owns the 
capability of reconstructing objects from incomplete and noisy data to make 
a spatial analysis upon the flaring quasar. We introduce the instrument and
the analysis method in Sect.2 and give our results in 
Sect.3. The results show clearly the detection of PKS0528+134 in the energy
range 3 MeV to 30 MeV. We also compare the intensity 
ratios and the light curves 
with those derived from EGRET and give our flaring
spectrum. In Sect.4, we give a discussion and draw the conclusion that the 
quasar kept on flaring in the energy range 3 MeV to 30 MeV during 
COMPTEL VP 213.  

\section{Instrument and analysis method}

COMPTEL is one of the four instruments aboard {\sl CGRO} and is sensitive
to $\gamma$-rays in the energy range 0.75 MeV to 30 MeV.
COMPTEL consists of two detector arrays. An incident $\gamma$-ray photon
is first Compton scattered in the upper detector and then interacts with the lower detector. Two orthogonal coordinates
($\chi$,$\psi$) which describe the direction of the scattered $\gamma$-ray
can be obtained from the positions of interaction occurring
in the two detectors and the Compton scattering  angle $\bar{\varphi}$ in the upper
detector can
be calculated from the measured energy deposits of the photon in the two detectors.
The incident direction
of the photon is then known to lie on a projected circle (with its center at 
($\chi$,$\psi$) and with radius $\bar{\varphi}$, called 'event circle')
on the sky for the ideal case of total
absorption in the lower detector. The scattering direction ($\chi$,$\psi$)
and the Compton scattering angle $\bar{\varphi}$ constitute a three-dimensional
data space in
which the spatial response of the instrument is cone-shaped. For the details
about COMPTEL see Sch\"{o}nfelder et al. (1993).

Inversion methods are needed for source reconstruction in three-dimensional
data space and a direct demodulation method is chosen by us. 
The direct demodulation method has already been successfully
applied to imaging analysis of scan observation data of slat collimator
telescopes, e.g.
the scan imaging of CygX-1 by the balloon-borne hard X-ray collimated
telescope HAPI-4 (Lu Z.G. et al. 1995)
and reanalyzing the 
EXOSAT-ME galactic plane survey (Lu F.J. et al. 1996).
It can also be applied to analyze observational data from a coded-mask aperture
telescope (Li 1995), rotating modulation telescope (Chen et al. 1998) and
Compton telescope (Zhang et al. 1997).

The principle of
direct demodulation (Li \& Wu 1994) is to
perform a deconvolution from the following correlation equation 
under proper physical constraints:
\begin{equation}
\label{ }
P'f=c
\end{equation}
where $P'=P^T P$, $c=P^T d$, $P$ is the point spread function matrix of 
COMPTEL and
$P^T$ its transpose, $d$ is the observational data and $f$ the
intensity distribution of the unknown sky and background.
The physical constraints
can be the upper and lower limits of intensity.
The process of direct demodulation 
consists of two main steps. The diffuse background
is derived first by iteratively solving the correlation equation under 
% continuous
constraints of continuity. Then, the intensity distribution of the object sky is obtained 
by solving
the correlation equation again under constraints of the produced diffuse
background as a lower limit. After subtracting the diffuse background, 
the true intensity distribution of the sources in the object
sky and the non-uniform component of the  background are
left. In this work iteration calculations are
carried out by using the Gauss-Seidel algorithm and error estimates derived
by the bootstrap technique.

\section{Results}

>From the beginning of the all-sky survey of COMPTEL PKS0528+134 has been
in the field of view several times. During observation VP 213 (March 23
-- 29, 1993) the source was located about 9.0$^{\circ}$ off the COMPTEL
pointing direction. With the direct demodulation method, we have analyzed the 
data from VP 213 in the four standard COMPTEL energy bands. 

Images of PKS0528+134 were obtained only in the 3-10 MeV and 
10-30 MeV bands (see Fig. 1a and Fig. 1b). In the 0.75-1 MeV and 1-3 MeV bands,
no convincing excess around the position of PKS0528+134 was 
found. Fig. 1a and Fig. 1b
show clearly the existence of a source at the position of PKS0528+134. 
In addition to the quasar, the Crab 
nebula and pulsar Crab also exist in the two skymaps and these two $\gamma$-ray
sources are resolved exactly. The flux density values of PKS0528+134 are found to be 
(16.6$\pm$5.95)$\times$10$^{-5}$ ph cm$^{-2}$ s$^{-1}$ in the 3-10 MeV band
and (7.71$\pm$3.03)$\times$10$^{-5}$ ph cm$^{-2}$ s$^{-1}$ in the 10-30 MeV 
band. The relative statistical significance is about 2.79$\sigma$ and
2.54$\sigma$. The flux values of the Crab in the two skymaps are 
(3.57$\pm$0.68)$\times$10$^{-4}$ ph cm$^{-2}$ s$^{-1}$ in Fig. 1a and
(9.57$\pm$2.57)$\times$10$^{-5}$ ph cm$^{-2}$ s$^{-1}$ in Fig. 1b. 
Therefore, although the quasar
became the brightest $\gamma$-ray source during its flare at energies above 
100 MeV, it was still weaker than the Crab at MeV energies.
The flux values  of PKS0528+134 obtained 
by Collmar et al. (1997) are 
(15.2$\pm$4.7)$\times$10$^{-5}$ ph cm$^{-2}$ s$^{-1}$ in the 
3-10 MeV band and
(3.3$\pm$1.7)$\times$10$^{-5}$ ph cm$^{-2}$ s$^{-1}$ in the  
10-30 MeV band. Our 3-10 MeV flux is close to their result and 10-30 MeV
flux is about 2.3 times higher. Both the 10-30 MeV fluxes, nevertheless, are
consistent with each other within the error bars.
Collmar et al. (1997) also gave an upper 
limit of
the 0.75-1 MeV flux and a very marginal 1-3 MeV flux, which indicates that 
PKS0528+134 was very weak and could hardly be detected 
at energies below 3 MeV. Our  
non-detection of PKS0528+134 in the 0.75-1 MeV and 1-3 MeV bands 
confirm this indication.
The other structures in both the skymaps reflect the non-uniform 
background from instrument
and the statistical fluctuations of the observational data.

We list in Table 1 the fluxes detected by EGRET and by COMPTEL at 
energies above 3 MeV
during three high states of PKS0528+134. Fluxes marked (*) are our 
results and fluxes of PKS0528+134 for COMPTEL VP 0 and VP 1 are from
Collmar et al. (1997). All the EGRET results in Table 1 are from
Mukherjee et al. (1996). With these fluxes, we have calculated the 
intensity ratios for every two
high states observed by EGRET and by COMPTEL respectively, and the results
are listed in Table 2. We can see 
from Table 2 that the 10-30 MeV intensity ratios are close
to those detected by EGRET at energies above 100 MeV and the intensity 
ratios of VP 213/VP 1 and VP 213/VP 0 
in the 3-10 MeV band are also high, which suggest that the 
flaring of PKS0528+134 was in phase in both the energy intervals 3-30 MeV and 
$\geq$ 100 MeV.
 The non-detections in the lower energy bands show its cessation.
The COMPTEL 3-10 MeV and 10-30 MeV light curves
together with the EGRET one (Mukherjee et al. 1996) are shown in Fig.2.
In Fig.2 our fluxes of PKS0528+134 for VP 213 are used and all other data 
points in the COMPTEL 3-10 MeV and 10-30 MeV light curves
are from Collmar et al. (1997). From Fig.2 one can see that the COMPTEL 
3-10 MeV and 10-30 MeV energy bands follow well the intensity trend as 
observed by EGRET above 100 MeV. 
  
The COMPTEL spectral results together with the simultaneously measured EGRET
spectra of PKS0528+134 for VP 1 and VP 213 are shown in Fig. 3a 
and Fig. 3b. The spectrum of the quasar in Fig. 3a is from Collmar et al. 
(1993b). The spectrum at energies above 30 MeV in Fig. 3b is from 
Mukherjee et al. (1996) and the other two data points in this figure are 
our results. The COMPTEL spectral results shown in Fig.3b 
are consistent with a simple extrapolation of the measured EGRET spectrum.
The spectral break in the 10-30 MeV band for COMPTEL 
VP 1 (see Fig. 3a) does not appear in the flaring spectrum
for VP 213 (see Fig. 3b) even down to the energy of 
3 MeV. The non-detection of PKS0528+134 at energies below 3 MeV
implies that a spectral break may occur around 3 MeV in Fig. 3b.
Such a spectral break is also required by the results of Collmar et al. (1997).

\section{Discussion and conclusion}

We have carried out  an imaging analysis upon the quasar PKS0528+134 during its flare
with the direct demodulation method. All our results agree with those given by
Collmar et al. (1997) except for a difference in the 10-30 MeV flux.
The 3-10 MeV flux obtained 
by us is close to that by Collmar et al. (1997) and our 10-30 MeV flux is
about a factor of 2.3 higher. Taking into account, however, the significance 
of the detections
and the errors of the derived fluxes, one can see that the two values for the 10-30
 MeV flux are in agreement.
The flux difference may result from the different
treatment of the background when analyzing the data. In our analysis, 
we search for 
the diffuse background only from the data and use them as lower limit
when solving again the modulation equation.
The non-uniform instrument  
background and the intensity distribution of the real 
sources are left
in the final map. In this way, we need no longer  consider the background prior to estimating intensities of the sources because both of them 
can be derived from the data simultaneously.  
We note that there is some structure near PKS0528+134 in Fig. 1b and 
no other structures around PKS0528+134 in Fig. 1a, 
which imply that 
the background around PKS0528+134 is relatively clean in the 3-10 MeV band 
but dirty and noisy in the 10-30 MeV band. The complicated 
background in Fig. 1b may therefore lead to the difference in the 10-30 MeV flux 
when using different analysis methods.
      
Although PKS0528+134 has been fully studied by Collmar et al. (1997), some 
important features for the flaring quasar are found by us in our reanalyzing 
COMPTEL data of VP 213 with the direct demodulation method.
The flaring quasar PKS0528+134 was detected with high luminosity
in the energy range 3 MeV to 30 MeV and not detected in the lower
COMPTEL energy bands; The intensity ratios
and the light curves measured by COMPTEL at energies above 3 MeV are 
consistent with those by EGRET; A spectral break at energies around
3 MeV is required from the combined spectrum measured simultaneously
by COMPTEL and EGRET during VP 213.
These features suggest that
PKS0528+134 still kept on flaring in the 3-30 MeV band at 
similar level as measured by EGRET above 100 MeV and no flaring below 3 MeV.
The inverse Comptonization (IC) models, which can be subdivided into 
the self generated synchrotron Compton model 
(SSC model; e.g. Maraschi et al. 1992, Bloom \& Marscher 1993) and the external 
radiation Compton model
(ERC model; e.g. Dermer \& Schlickeiser 1993, Sikora et al. 1994, 
Blandford \& Levinson 1995) according to the origin of low-energy photons
scattered to $\gamma$-energies by the relativistic jet electrons, 
are generally used to explain the production 
of high energy $\gamma$-rays from quasars.
To explain the spectral break for the flaring spectrum of PKS0528+134, however, 
we need a combined ERC / SSC model (B\"{o}ttcher \& Collmar 1998) 
which introduces a high-energy electron pair population with 
a low-energy cutoff in its Lorentz-factor distribution 
(B\"{o}ttcher et al. 1997).
Our result for the flaring spectrum is consistent with this model. 
Our detections and non-detections of the flare at MeV
energies put hard constraints on the relative low-energy cutoff, thus  
helping to further understand the emission process that produces the flare.

\begin{acknowledgements}
We gratefully acknowledge
 Dr. W. Collmar for kindly providing all the data used in this work and
for many valuable comments. This work is supported by the National Natural
Science Foundation of China under grant 19673010.
\end{acknowledgements}

\onecolumn

\headsep=2cm
\Large
\begin{table}
\caption[]{Fluxes of PKS0528+134 measured by COMPTEL and EGRET in three
high states.}
\begin{tabular}{cccc}\hline
\multicolumn{1}{c}{VP}&\multicolumn{1}{c}{EGRET}&\multicolumn{2}{c}{COMPTEL}\\ 
 &$\geq$ 100 MeV&10-30 MeV&3-10 MeV \\
 &($\times10^{-7}$ph/cm$^{-2}$s$^{-1}$)&($\times10^{-5}$ph/cm$^{-2}$s$^{-1}$)
&($\times10^{-5}$ph/cm$^{-2}$s$^{-1}$) \\ \hline
0   &     12.9$\pm$0.9 &    3.2$\pm$1.0   &   4.4$\pm$2.5\\ 
1   &   8.5$\pm$0.8    &  3.0$\pm$1.0    &   $\leq$ 7.2\\ 
213   & 30.8$\pm$3.5   & 7.71$\pm$3.03$^{*}$   &   16.60$\pm$5.95$^{*}$\\ \hline
\end{tabular}
\vspace{1cm}
\end{table}

\begin{table}
\caption[]{Intensity ratios of PKS0528+134 measured by COMPTEL and EGRET.} 
\begin{tabular}{cccc}\hline
\multicolumn{1}{c}{Intensity ratio}&\multicolumn{1}{c}{EGRET}&\multicolumn{2}
{c}{COMPTEL}\\ 
   &   $\geq$ 100 MeV  &  10-30 MeV  &  3-10 MeV\\ \hline
VP 1/VP 0  &  0.66$\pm$0.08   &  0.94$\pm$0.43  &  $\leq$ 1.64\\ 
VP 213/VP 0  &   2.39$\pm$0.32  &  2.41$\pm$1.21  &  3.77$\pm$2.53\\ 
VP 213/VP 1  &   3.62$\pm$0.54  &  2.57$\pm$1.32  &  $\geq$ 2.31 \\ \hline
\end{tabular}
\end{table}

\begin{figure}[h]
\vbox to4.in{\rule{0pt}{4.in}}
\includegraphics{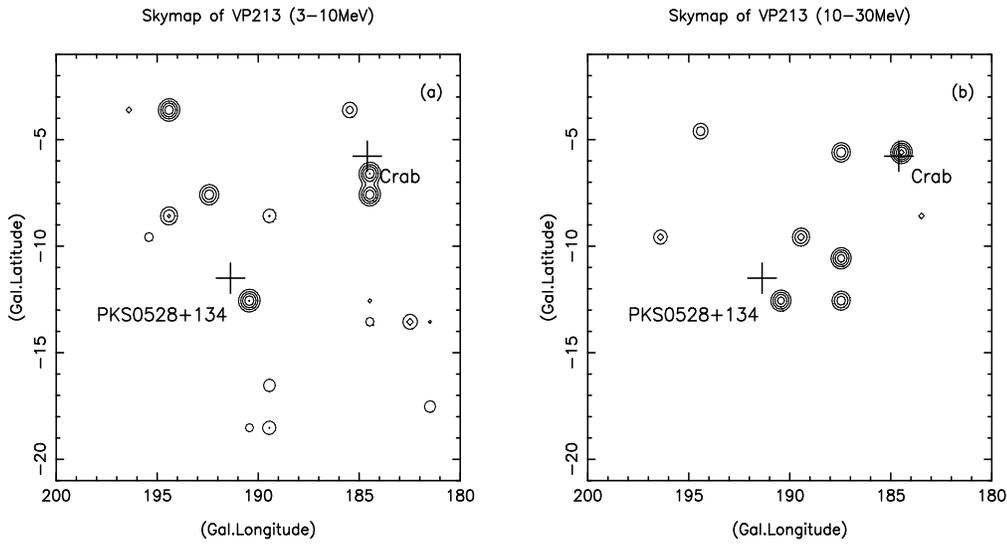}
\caption{$\bf{a}$ The direct demodulation map of VP 213 in the 3-10 MeV band. 
$\bf{b}$ The direct demodulation map of VP 213 in the 10-30 MeV band.
Apart from Crab and PKS0528+134, the other source features emerged in these two
maps are generally insignificant and most of their statistical 
significances are below 2$\sigma$.}
\end{figure}

\newpage

\begin{figure}[h]
\vbox to4.in{\rule{0pt}{4.in}}
\includegraphics{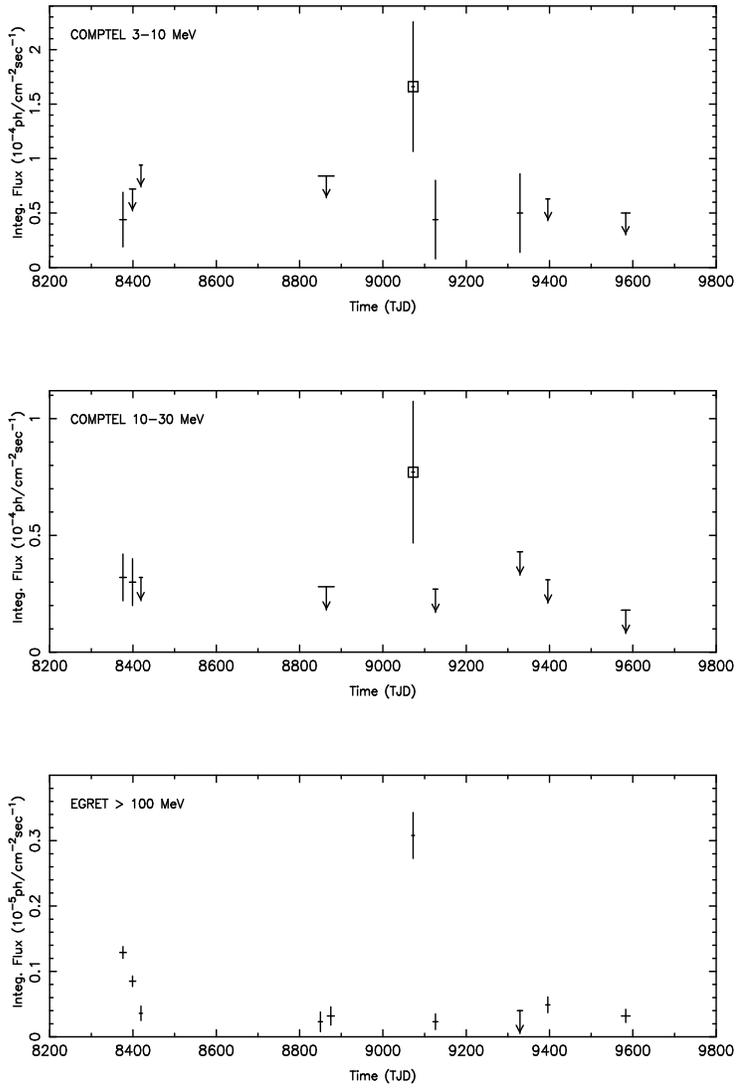}
\caption{Time history of PKS0528+134 as measured by COMPTEL in the 3-10 MeV and 
10-30 MeV bands and by EGRET above 100 MeV (Mukherjee et al. 1996). Data points
with square symbol are our results and all other COMPTEL data points are from
Collmar et al. (1997). The error bars are 1$\sigma$ and the upper limits are
2$\sigma$.}
\end{figure}

\headsep 2.5in
\begin{figure}[h]
\vbox to4.in{\rule{0pt}{4.in}}
\includegraphics{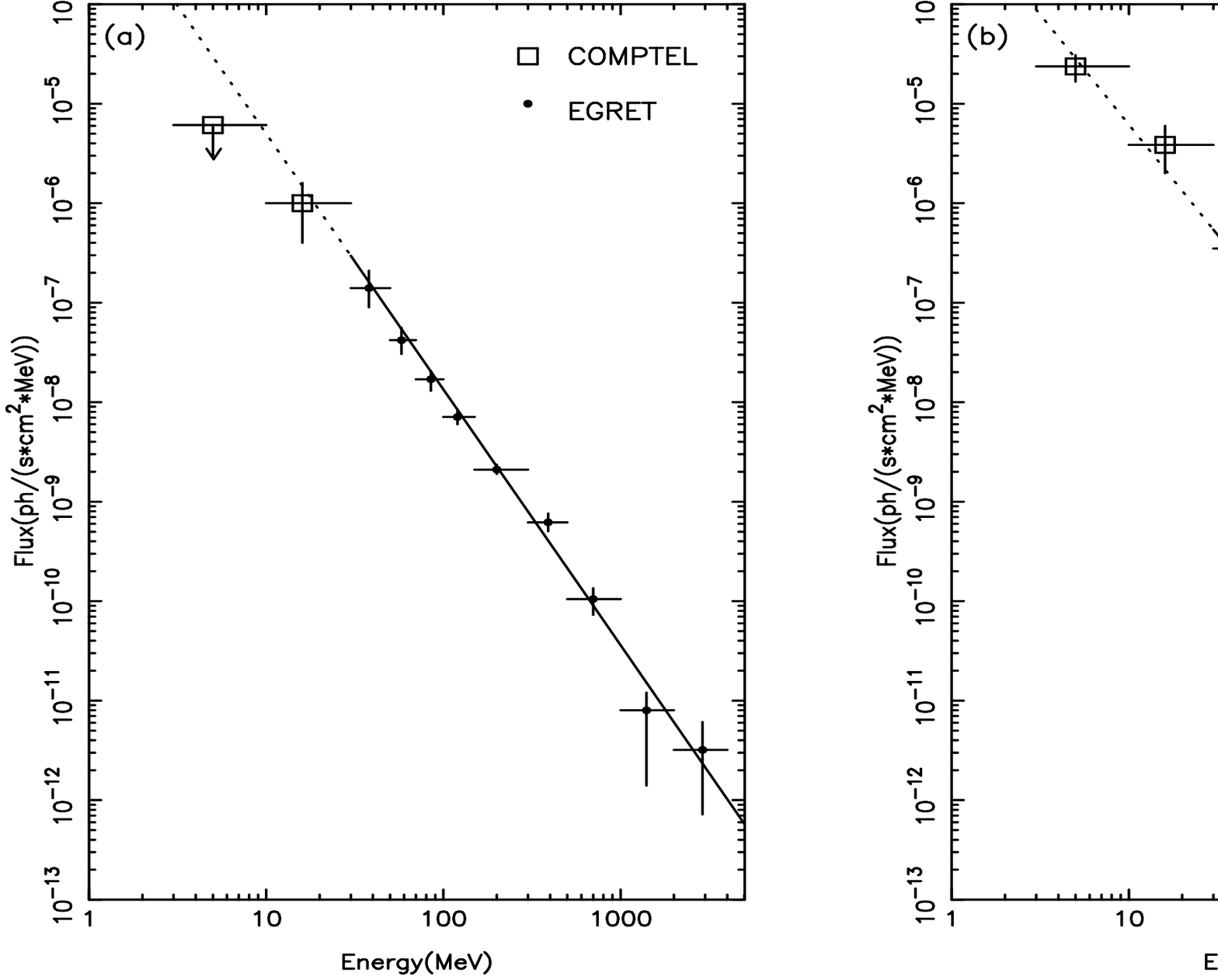}
\caption{$\bf{a}$ Energy spectrum of PKS0528+134 measured by COMPTEL and EGRET 
during VP 1.
$\bf{b}$ Energy spectrum of PKS0528+134 measured by COMPTEL and 
EGRET during VP 213.}
\end{figure}


\begin{thebibliography}{}
\bibitem[1995]{blandford} Blandford R., Levinson A., 1995, ApJ 441, 79
\bibitem[1993]{bloom} Bloom S.D., Marscher A.P., 1993, AIP Conference 
Proceedings 280, 578
\bibitem[1998]{bottcher} B\"{o}ttcher M., Collmar W., 1998, A\&A 329, L57
\bibitem[1997]{bottcher} B\"{o}ttcher M., Mause H., Schlickeiser R.,  
1997, A\&A 324, 395  
\bibitem[1985]{bregman} Bregman J.N., Glassgold A.E., Huggins P.J., Kinney A.L.,
1985, ApJ 291, 505
\bibitem[1998]{chen} Chen Y., Li T.P., Wu M., 1998, A\&AS 128, 363
\bibitem[1993a]{collmar} Collmar W., Diehl R., Lichti G.G., et al., 
1993a, AIP Conference Proceedings 280, 483 
\bibitem[1993]{collmar} Collmar W., Bloemen H., Bennett K., et al.,
1993b, Proc. of the 23rd ICRC conference, Vol.1, 168, Calgary, Canada
\bibitem[1998]{collmar} Collmar W., Bennett K., Bloemen H., et al., 1997,
A\&A 328, 33
\bibitem[1993]{dermer} Dermer C.D., Schlickeiser R., 1993, ApJ 416, 458
\bibitem[1988]{fugmann} Fugmann W., Meisenheimer K., 1988, A\&AS 76, 145
\bibitem[1993]{hunter} Hunter S.D., Bertsch D.L., Dingus B.L., et al.,
1993, ApJ 409, 134
\bibitem[1995]{li} Li T.P., 1995, Exper. Astron. 6, 63
\bibitem[1994]{li} Li T.P., Wu M., 1994, Ap\&SS 215, 213
\bibitem[1996]{lu} Lu F.J., Li T.P., Sun X.J., Wu M., Page C.G., 1996,
A\&AS 115, 395
\bibitem[1995]{lu} Lu Z.G., Wang J.Z., Li Y.G., Shen P.R., 1995, Nucl. Instr. 
Meth. A 362, 551
\bibitem[1992]{maraschi} Maraschi L., Ghisellini G., Celotti A., 1992, 
ApJ 397, L5
\bibitem[1996]{mukherjee} Mukherjee R., Dingus B.L., Gear W.K., et al.,
1996, ApJ 470, 831
\bibitem[1993]{nolan} Nolan P.L., et al., 1993, IAU circular 5802
\bibitem[1993]{schonfelder} Sch\"{o}nfelder V., Aarts H., Bennett K., et al.,
 1993, ApJS 86, 657
\bibitem[1994]{sikora} Sikora M., Begelman M.C., Mitchel C., et al., 1994, 
ApJ 421, 153
\bibitem[1997]{zhang}Zhang S., Li T.P., Wu M., et al., 1997, AIP Conference
Proceedings 410, 578

\end{thebibliography}
\end{document}